\documentclass[aps,prl,twocolumn,unsortedaddress]{revtex4-1}

\usepackage{graphicx}

\begin{document}


\title{Multi-channel exchange-scattering spin polarimetry}


\author{Fuhao Ji$^{2}$, Tan Shi$^{2}$, Mao Ye$^{1}$, Weishi Wan$^{3}$, Zhen Liu$^{2}$, Jiajia Wang$^{4}$, Tao Xu$^{4}$, and Shan Qiao$^{1, 4, }$}
\email{qiaoshan@mail.sim.ac.cn; }
\affiliation{$^{1}$State Key Laboratory of Functional Materials for Informatics,
Shanghai Institute of Microsystem and Information Technology,
Chinese Academy of Sciences, 865 Changning Road,
Shanghai 200050, China \\
$^{2}$  Physics Department, Laboratory of advanced Materials, and Surface Physics Laboratory (National Key Laboratory), Fudan University, 2005 songhu road, Shanghai, 200438, China\\
$^{3}$ Advanced Light Source, Lawrence Berkeley National Laboratory, 1 Cyclotron Road, Berkeley, CA 94720, USA \\
$^{4}$ School of physical science and technology, ShanghaiTech University, 319 Yueyang Road, Shanghai, 200031, China \\}


\date{\today}

\begin{abstract}
Electron spin takes critical role in almost all novel phenomena discovered in modern condensed matter physics (High-temperature superconductivity , Kondo effect, Giant Magnetoresistance, topological insulator, quantum anomalous Hall effect, etc.). However, the measurements for electron spin is of poor quality which blocks the development of material sciences because of the low efficiency of spin polarimeter. Here we show an imaging type exchange-scattering spin polarimeter with 5 orders more efficiency compared with a classical Mott polarimeter. As a demonstration, the fine spin structure of electronic states in bismuth (111) is investigated, showing the strong Rashba type spin splitting behavior in both bulk and surface states. This improvement pave the way to study novel spin related phenomena with unprecedented accuracy.
\end{abstract}


\maketitle

Almost all novel phenomena discovered in modern condensed matter physics, for example, High-temperature superconductivity \cite{ZX}\cite{stewart}, Kondo effect, Giant Magnetoresistance \cite{baibich}, topological insulator \cite{Hasan}, quantum anomalous Hall effect \cite{Chang}, are related with exchange or spin-orbital interactions both strongly related with the spin of electrons. However, since the original work by Mott, the measurement of electron spin remains a challenge. The efficiency of spin polarimeters \cite{Kirschner}\cite{Scheinfein}\cite{Dunning}\cite{Gay}\cite{Qiao} based on spin-orbital interaction (SOI) is about $10^{-4}$, that it takes ten thousand times of data acquisition time to obtain a spin-resolved spectrum with the same statistical error compared with regular photoelectron measurements. Recently, very low energy electron diffraction (VLEED) type\cite{Okuda} spin polarimeter has been developed employing the exchange interaction between incident polarized electrons and those in the ferromagnetic targets\cite{Bertacco}\cite{BertaccoRSI}, where the efficiency is greatly improved up to about $10^{-2}$, being hundred times higher than that based on SOI. On the other hand, the state-of-art analyzer for angle-resolved photoelectron spectroscopy (ARPES) developed by K. Siegbahn group enables the ARPES measurements in multi-channel, or image-type mode. With this technique, the electrons with the same emission angle (i.e. the same momentum) are focused by the electron optics to a certain positions at the entrance slit and disperse along another perpendicular axis when pass the hemispherical energy analyzer (HEA).  As a result, the intensity recorded at the exit panel of HEA directly visualizes the band structures in a defined energy-momentum (E-k) space with more than $10^4$ channels being observed simultaneously. However, the situation becomes totally different in the field of spin-resolved ARPES (SARPES). The measurement could be only conducted for one specific point in the E-k space with certain emission angle and kinetic energy at once because of the lack of image type spin polarimeter.  Even with VLEED-type spin polarimeter of improved efficiency, a typical SARPES measurement takes days and provides only a few representative spectra, sacrificing the detailed spin-polarized information with compromised energy resolution. The spin-polarized low-energy electron diffraction (SPLEED)\cite{Kirschner}, which is also based on SOI mechanism, has been utilized as spin-filtering mirrors for image type spin measurements and a total of $10^3$ channels has been demonstrated \cite{Kolble}\cite{Kutnyakhov}\cite{Vasilyev}. In this intelligent design, the electron optics makes a virtual image behind the scattering target and the final image forms on the detector plane by specular reflection. To challenge the ultimate limit of the SARPES efficiency, multichannel spin polarimeter based on exchange interaction is indispensable and which cannot be simply achieved by only the change of non-ferromagnetic scattering target to ferromagnetic one in the specular reflection design discussed above because of the Stoner excitation\cite{KirschnerStoner} which will change the momentum of electrons and destroy the function of the electron optics. To avoid this problem, the formation of real image on ferromagnetic scattering target is necessary.

\begin{figure*}
\includegraphics[width=.75\textwidth]{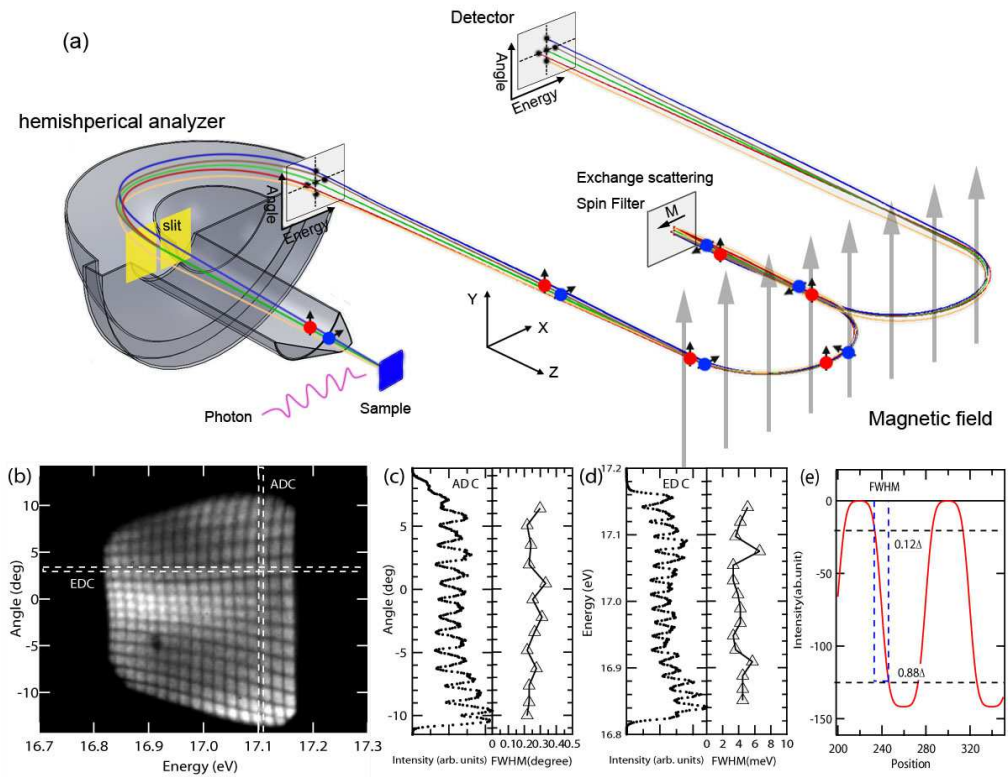}
\caption{\label{figure1} (a) A schematic view of the multichannel spin polarimeter attached to an energy analyzer. (b) An uncorrected electron intensity image on the 2D detector after performing a 6 eV VLEED reflection, the dark lines originates from a mesh set at the entrance of the spin polarimeter. (c) and (d) the intensity distribution cut from (b) along angle and energy directions, the instrumental FWHMs of the multichannel spin polarimeter are obtained by the method illustrated in (e).}
\end{figure*}

In this letter, we report the first realization of a multichannel exchange scattering spin polarimeter, which gives 6786 independent channels with average efficiency up to $8.5\times10^{-3}$ for each single channel. Such dramatic improvement results in about $5\times10^5$ times more efficiency than a classical Mott polarimeter. The ultrahigh efficiency enables the observation of the detailed spin-polarized electronic structures demonstrated by the measurements on bismuth (111) with unprecedentedly high resolution, where the detailed momentum dependence of the spin splitting magnitude is quantitatively provided, showing a strong Rashba type splitting behavior in the surface scattered bulk states as well as the surface band.

\begin{figure*}
\includegraphics[width=.8\textwidth]{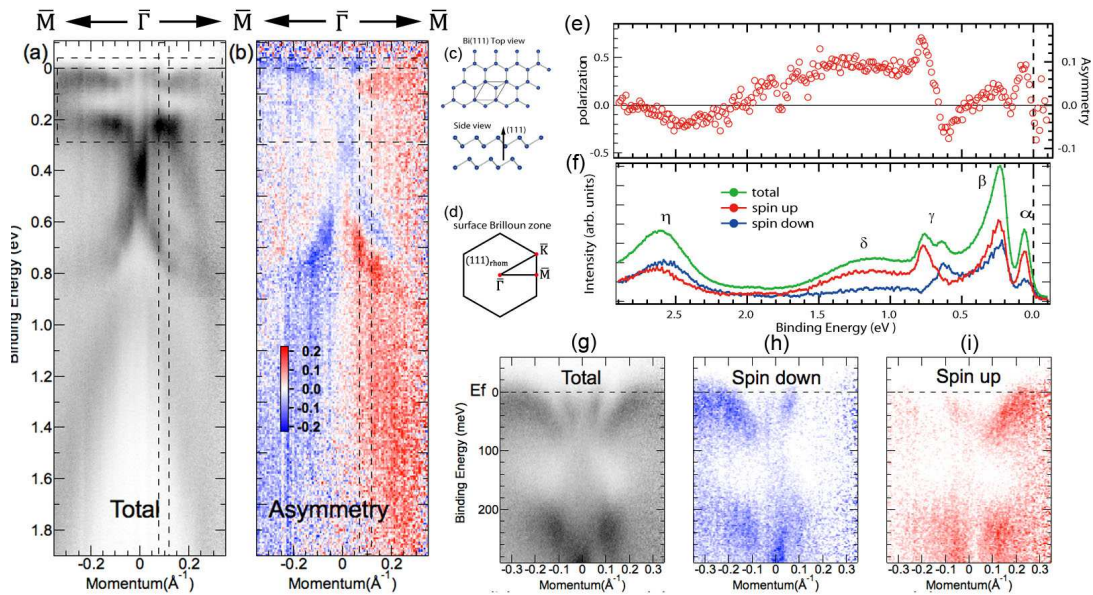}
\caption{\label{figure2} (a) The spin integrated ($I_{total}=(I_++I_-)/2$) band dispersion along $\bar{M}-\bar{\Gamma}-\bar{M}$ direction of Bi(111) surface obtained by the multichannel spin polarimeter. (b) The scattering asymmetry obtained from two spectrums taken under opposite target magnetization $A=(I_+-I_-)/(I_++I_-)$. (c) and (d) A schematic view of the crystal structure and surface Brillioun zone of Bi(111). (e) and (f) The spin polarization and intensity EDC of marked area in Figure2(a), the corresponding wave vector is $K_{\parallel} = 0.1\AA^{-1}$. The strongest peak of spin polarization at Binding energy 0.8 eV is used to calibrate the effective Sherman function of the spin polarimeter. (g), (h), (i) The spin integrated and resolved spectrum of Rashba state near Fermi surface, corresponding to the upper marked area in figure 2(a). Spin up and down spectrum is obtained from $I_{\uparrow \downarrow}= (1\pm P)I_{total}/2$. Bands with opposite spin separate along $\pm K$, showing the reflection symmetry due to TRI.}
\end{figure*}

The Sherman function $S_{eff}$ that defined as the ratio between observed scattering asymmetry and the intrinsic spin-polarization of incident electrons, significantly effects the efficiency of spin polarimeter, or figure of merit ($FOM$), which is $\frac{I}{I_0}\times S_{eff}^2$, where $I$ and $I_0$ are the intensity of scattered and incident electrons, respectively. In order to maximize the efficiency of spin measurements using oxygen passivated iron film as the exchange scattering target, the trajectory of the incident electron has to be normal to the target surface \cite{Bertacco}. The key idea of realizing a multi-channel VLEED (MCVLEED) type spin polarimeter is to use a homogeneous magnetic field as a beam deflector, which can separate the incident and reflected electrons. Figure~\ref{figure1}(a) shows a schematic view of the MCVLEED spin polarimeter. After passing through the VG Scienta R3000 HEA, electrons are separated by their energies and take off angles and form a 2D image of band dispersion on the exit plane. Electron lens system has been carefully designed to transfer the electron beam image from the exit plane to the ferromagnetic $Fe(001)-P(1\times1)-O$ target and from the target to the two dimensional electron detector composed of two microchannel plates (MCP) and a fluorescence screen. Here, the time reversal anti-symmetry of magnetic field plays the key role: First, the electron beam turns 180 degrees and enters the VLEED spin detection system, ensuring a normal incident reflection; Then the reflected electrons re-enter the magnetic field and turns another 180 degrees again, being separated from the incident electron beam and turned to the opposite direction. The analysis shows that the Larmor precession of electron spin in the magnetic field does not affect the measurements \cite{Supp}. Ray tracing results of the electron optics using SIMION program \cite{simion} give more than 100 distinguishable channels for both energy and angular directions, which means in principle $10^4$ data points can be acquired simultaneously when performing the VLEED measurements \cite{CPC}. Another advantage of this design is that the switching between spin-resolved and spin-integrated modes becomes quite simple by decreasing the magnetic field to a half value , in which the electron beam will turn 180 degrees with two times larger radius and directly collected by the 2D electron detector with even higher spatial resolution.

In order to examine the spatial resolution of the electron optics, which is the key factor to determine the number of distinguishable channels, a mesh is set at the exit plane of HEA. The width of the mesh strips is 0.2 mm and the distance between strips is 1 mm. The size of the imaging region is 20 mm$\times$20 mm, corresponding to an acceptance angle of 20 degree and an energy window of 375 meV.  Figure~\ref{figure1}(b) shows the electron intensity image recorded by the 2D detector after a practical 6 eV VLEED reflection with the entrance slit of HEA been illuminated by a hot cathode electron gun. Angular and energy distribution curves are extracted to determine the spatial resolution and number of distinguishable channels, as shown in Figure~\ref{figure1}(c) and (d). The spatial resolution, i.e., FWHM can be estimated from the broadening of peak edges. FWHM should be smaller than the width in which the edge height changes from $12\%$ to $88\%$ (Figure~\ref{figure1}(e)). In angular (energy) direction, an average FWHM of 0.255 degree (4.31 meV) gives 78 (87) distinguishable channels. This result shows the possibility to achieve at least $78\times87 = 6786$ distinguishable channels. Here, the number of distinguishable channels is mainly limited by the electron optical aberrations. Because of the aberration of dipole magnetic deflector, the square object region transfers into a trapezoid form. Nevertheless, the distortion does not affect resolutions and can be corrected during post processing.

To demonstrate the ultra-high efficiency and quantitatively determine the performance of MCVLEED, we performed the SARPES measurements on single crystal Bi(111) thin film \cite{Hirahara}, which is reported to show complex spin-polarized electronic structures\cite{Kimura}. Previous studies have confirmed a Rashba type spin splitting band existing on its surface, which is induced by a broken space inversion symmetry as well as a strong SOI\cite{Kimura}\cite{Koroteev}\cite{HiraharaSpin}\cite{OkudaSpin}. $Fe(001)-P(1\times1)-O$ target is prepared following ref. \cite{Okuda} in a ultra-high vacuum (UHV) environment. The quality of the single crystal target is checked by low energy electron diffraction (LEED). High quality Bi(111) thin film is grown at room temperature on a $Si(111) 7\times7$ substrate cleaned in situ by resistive heat treatments\cite{Hirahara}. Figure~\ref{figure2}(a) shows the band dispersion of Bi(111) surface along ¦£$\bar{M}-\bar{\Gamma}-\bar{M}$ direction excited by a He I$\alpha$ emission line ($h\nu = 21.2eV$) at T = 80 K and Figure~\ref{figure2}(b) shows the corresponding asymmetry in the same E-k region. The spin polarization is then determined as $P = A/S_{eff}$, where A is the asymmetry shown in Figure~\ref{figure2}(b).  Two sets of typical Rashba-type spin-splitting bands, located near Fermi energy and binding energy ($E_B$) around 0.5 - 0.8 eV, have been clearly resolved in Figure~\ref{figure2}(b), which are originated from the surface states and bulk continuum states, respectively\cite{Hirahara}\cite{Kimura}.  Spin resolved energy distribution curves (EDC) and momentum distribution curves (MDC) can be extracted from the 2D spectra with fixed $K_\parallel$ or energy. In order to obtain $S_{eff}$, the asymmetry at $K_{\parallel} = 0.1\AA^{-1}$ as a function of binding energy was extracted (Figure~\ref{figure2}(e)) and compared with the reported polarization \cite{OkudaSpin}, where the peak at $E_B = 0.8 eV$ is chosen to perform calibration. For a typical measurement with an energy window of 375meV, the optimal center scattering energy is found to be 5.8 eV \cite{Supp}, which gives an average $<S_{eff}> = 0.225$ and the corresponding $<FOM> = 8.5\times10^{-3}$. Combining with the number of distinguishable channels, we derive the "two-dimensional figure of merit"\cite{Kolble} $FOM_{2D}=N\times<FOM> = 57.6$, which is about $5\times10^5$ times more efficient than a classical Mott type single channel spin polarimeter.

\begin{figure}
\includegraphics[width=8cm]{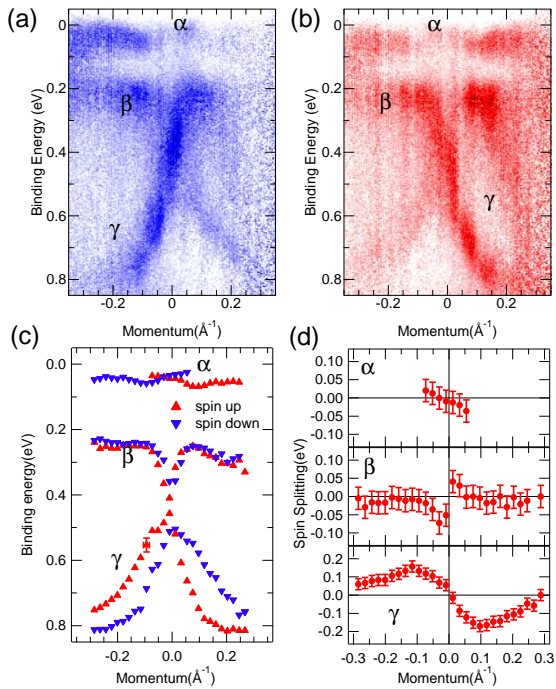}
\caption{\label{figure3} (a) and (b) The spin down and up band dispersion of both surface Rashba state and bulk states. (c) The fitting Lorentzian peak positions in k space of three bands marked in Figure3 (a) and (b), spin up and down band dispersions are put together. The uncertainty of the peak position are illustrated by error bar resulting from energy resolution. (d) The momentum dependent spin splitting magnitude (spm) obtained from (c), $E_{Spm}(n,k)=E_{\uparrow}(n,k)- E_{\downarrow}(n,k)$.}
\end{figure}

Such high efficiency results in significantly shortened acquisition time (5h for a spin resolved ARPES measurement as shown in Figure~\ref{figure2}(a) and (b)). And more importantly, the spin resolved band dispersion in the momentum space can be obtained with unprecedentedly high resolution, which provides the possibility of investigating the fine spin structure of electronic states in a wide momentum range. In spin integrated spectrum, one can easily see the Rashba spin splitting band near the Fermi surface, showing the petal like electron and hole pockets at binding energy $E_B = 50 meV$ (Figure~\ref{figure2}(g)). Spin resolved band dispersion can be obtained with the formula $I_{\uparrow \downarrow}= (1\pm P)I_{total}/2$, as shown in Figure~\ref{figure2}(h) and (i). Remarkably, the surface Rashba bands can be clearly resolved as the combination of one parabolic surface state and its Kramers counterpart. The reflection symmetry of two spectra with opposite spin originates from TRI of spin-orbital interaction. Figure~\ref{figure3}(b) ((c)) shows the spin down (up) distribution of both bulk and surface electronic states. The double band structure "$\gamma$" at $E_B = 0.7eV$ is separated into two single bands with opposite spin. By fitting the spin up and down EDCs with Lorentzian peaks, the spin resolved band dispersion can be obtained quantitatively. Figure 3(d) shows the fitting peak positions of three bands marked in Figure~\ref{figure3}(b) and (c), one can easily see the difference between bands with opposite spin. Energy difference between spin-up and spin-down peak positions gives the momentum dependent spin splitting magnitude in k space, as shown in Figure 3(d). For surface Rashba band denoted as "$\alpha$", the parabolic band dispersion shift toward $\pm k$ direction, showing a typical "Rashba type" linear spin splitting behavior around $\bar\Gamma$ point \cite{Koroteev}

$E_R =\pm a_R\times k$

The spin orbital coupling constant $a_R$ (Rashba parameter) is determined to be 0.54 eV$\AA$ by linear fitting the splitting magnitude, which is consistent with previous reports \cite{Ishizaka}. On the other hand, the polarized bulk states, denoted as "$\beta$" and "$\gamma$" show more complicated spin splitting behavior. Although the spin splitting is still antisymmetric about gamma bar point, the splitting magnitude deviated from linearity, where the splitting magnitude increase with $K_{\parallel}$ around the $\bar\Gamma$  point and then decrease eventually to zero. Interestingly, the surface scattered bulk state "$\gamma$" \cite{Kimura} shows even stronger spin splitting, and the splitting magnitude reaches its maximum at $K_{\parallel}=0.10\AA^{-1}$ as large as 160 meV. The coupling constant is determined to be 1.69 in the linear region, which is about 3 times that of the surface Rashba band.

We would like to emphasize that the acquiring time of above SARPES data is almost comparable with conventional ARPES measurement. The realization of multi-channel spin-polarimeter based on exchange scattering mechanism enables the quantitative data analysis on the details of spin-polarized electronic states, which have been widely applied to spin-integrated ones by conventional ARPES. This technique will significantly benefit the understanding of the mechanism of novel quantum materials with strong spin-orbital-coupling, such as the 2D and 3D topological insulators, Weyl semi- metal \cite{Weng}. Moreover, the MCVLEED spin-polarimeter is completely compatible with other state-of-art imaging type electron spectrometers like nanoAPRES and trARPES, which will enable the studies on spin-polarized electronic states of nano-structures and their dynamics.

\begin{acknowledgments}
We wish to express our gratitude to Prof. J. Kirschner in Max Planck Institute of Microstructure Physics for enlightening discussions on Stoner excitation. We also gratefully acknowledge Prof. T. Okuda in Hiroshima University for discussions on ferromagnetic target preparing. This work was supported by National Natural Science Foundation of China (Grants Nos. 10979021, 11027401, 11174054, 11304338 and 11227902), the Ministry of Science and Technology of China (National Basic Research Program Grant No. 2011CB921800), the ¡°Strategic Priority Research Program (B)¡± of the Chinese Academy of Sciences (Grant No. XDB04010100) and Helmholtz Association through the Virtual Institute for Topological Insulators (VITI).
\end{acknowledgments}


\end{document}